\documentclass[10pt,onecolumn,english,twocolumn]{IEEEtran}
\usepackage[T1]{fontenc}
\usepackage[latin9]{inputenc}
\usepackage{geometry}
\usepackage{amsmath}
\usepackage{stackrel}
\usepackage{graphicx}

\makeatletter

\providecommand{\tabularnewline}{\\}

\usepackage{babel}
\usepackage{algorithm}
\usepackage{algpseudocode}

\makeatother

\usepackage{babel}
\begin{document}
\title{Half-Marker Codes for Deletion Channels with Applications in DNA Storage}
\author{Javad Haghighat and Tolga M. Duman,~\IEEEmembership{Fellow,~IEEE}\thanks{J. Haghighat is with the Department of Electrical and Electronics
Engineering, TED University, 06420 Ankara, Turkey. T. M. Duman is
with the Department of Electrical and Electronics Engineering, Bilkent
University, 06800 Ankara, Turkey (e-mail: javad.haghighat@tedu.edu.tr,
duman@ee.bilkent.edu.tr)}\thanks{This work was funded by the European Union through the ERC Advanced
Grant 101054904: TRANCIDS. Views and opinions expressed are, however,
those of the authors only and do not necessarily reflect those of
the European Union or the European Research Council Executive Agency.
Neither the European Union nor the granting authority can be held
responsible for them. }}
\maketitle
\begin{abstract}
DNA data storage systems face significant challenges, including insertion,
deletion, and substitution (IDS) errors. Therefore, designing effective
synchronization codes, i.e., codes capable of correcting IDS errors,
is essential for DNA storage systems. Marker codes are a favorable
choice for this purpose. In this paper, we extend the notion of marker
codes by making the following key observation. Since each DNA base
is equivalent to a $2$-bit storage unit, one bit can be reserved
for synchronization, while the other is dedicated to data transmission.
Using this observation, we propose a new class of marker codes, which
we refer to as half-marker codes. We demonstrate that this extension
has the potential to significantly increase the mutual information
between the input sequence and the soft outputs of an IDS channel
modeling a DNA storage system. Specifically, through examples, we
show that when concatenated with an outer error-correcting code, half-marker
codes outperform standard marker codes and significantly reduce the
end-to-end bit error rate of the system. 
\end{abstract}


\section{\protect\label{sec:Introduction}Introduction}

DNA data storage systems offer a tremendous increase in recording
densities, as well as significantly longer lifetimes compared to conventional
storage systems \cite{Milenkovic2024},\cite{Shomorony2021}. The
cost of DNA storage systems has constantly reduced throughout the
last decade, thanks to the invention of cost-efficient DNA synthesis
(writing) and DNA sequencing (reading) techniques, including the nanopore
sequencing method \cite{Deamer2016}. However, such cost-efficient
techniques amplify the presence of insertion, deletion, and substitution
(IDS) errors \cite{Milenkovic2024}-\cite{Mao2018} . Unlike substitution
errors, which only alter one base, an insertion or deletion error
shifts all subsequent bases and can potentially corrupt large portions
of data. Hence, even a few insertion and deletion errors in a transmitted
block can render standard error-correcting codes ineffective. Therefore,
the synthesized data must be protected by codes designed to correct
insertion and deletion errors, also known as synchronization errors.
The design of efficient coding and decoding schemes for transmission
over channels with synchronization errors has been considered in several
recent works including \cite{Goshkoder2024}-\cite{Wang2024} and
the references therein, among others.

Marker codes, originally introduced in \cite{Sellers1962}, are a
class of synchronization error-correcting codes, which are broadly
employed in DNA storage systems \cite{Ma2024}, \cite{Welter2023},
\cite{Sriniva2021} due to their simple structure and efficient decoding
algorithms. In standard marker coding, the values of certain transmitted
symbols referred to as marker symbols, which are located at specified
positions of the transmitted sequence, are fixed, and known at the
decoder. The decoder exploits this \textit{a priori} information,
i.e., positions and values of the marker symbols, to correct the synchronization
errors by employing a forward-backward (FB) decoding algorithm \cite{Davey2001}.

In this paper, we introduce a new class of marker codes for DNA storage
systems, referred to as half-marker codes. A half-marker symbol is
defined as a symbol of a $4$-ary alphabet in which one bit has a
fixed value and contributes to maintaining synchronization, whereas
the second bit is an information-carrying bit that may take $0$ or
$1$ with equal probabilities (depending on the information bit embedded
in this symbol). Therefore, unlike standard marker symbols that are
designed solely for synchronization purposes, half-markers are synchronization
symbols that are also capable of carrying one bit of information.
We show that for a digital communication system that models a DNA
storage system, the mutual information between the input symbols and
their corresponding soft outputs may significantly increase when standard
marker codes are replaced by the proposed half-marker codes. In addition
to mutual information estimation, we also provide explicit concatenated
coding schemes when a low-density parity check (LDPC) code is applied
as the outer code, and the marker code is employed as the inner code.
Our numerical results show that the specific half-marker codes have
the potential to significantly reduce the end-to-end bit error rate
compared to standard marker codes in concatenated coding schemes.

The system model and preliminary definitions are given in Section
\ref{sec:System-Model}. In Section \ref{sec:Half-Marker-Codes:-Definition},
we present the proposed half-marker codes and discuss approaches to
evaluate their efficiency. In Section \ref{sec:Numerical-Results},
performance of the proposed half-marker codes is compared with the
performance of the standard markers via numerical examples. Section
\ref{sec:Conclusions} concludes this paper.

\section{\protect\label{sec:System-Model}System Model}

\begin{figure}
\centering\includegraphics[scale=0.25]{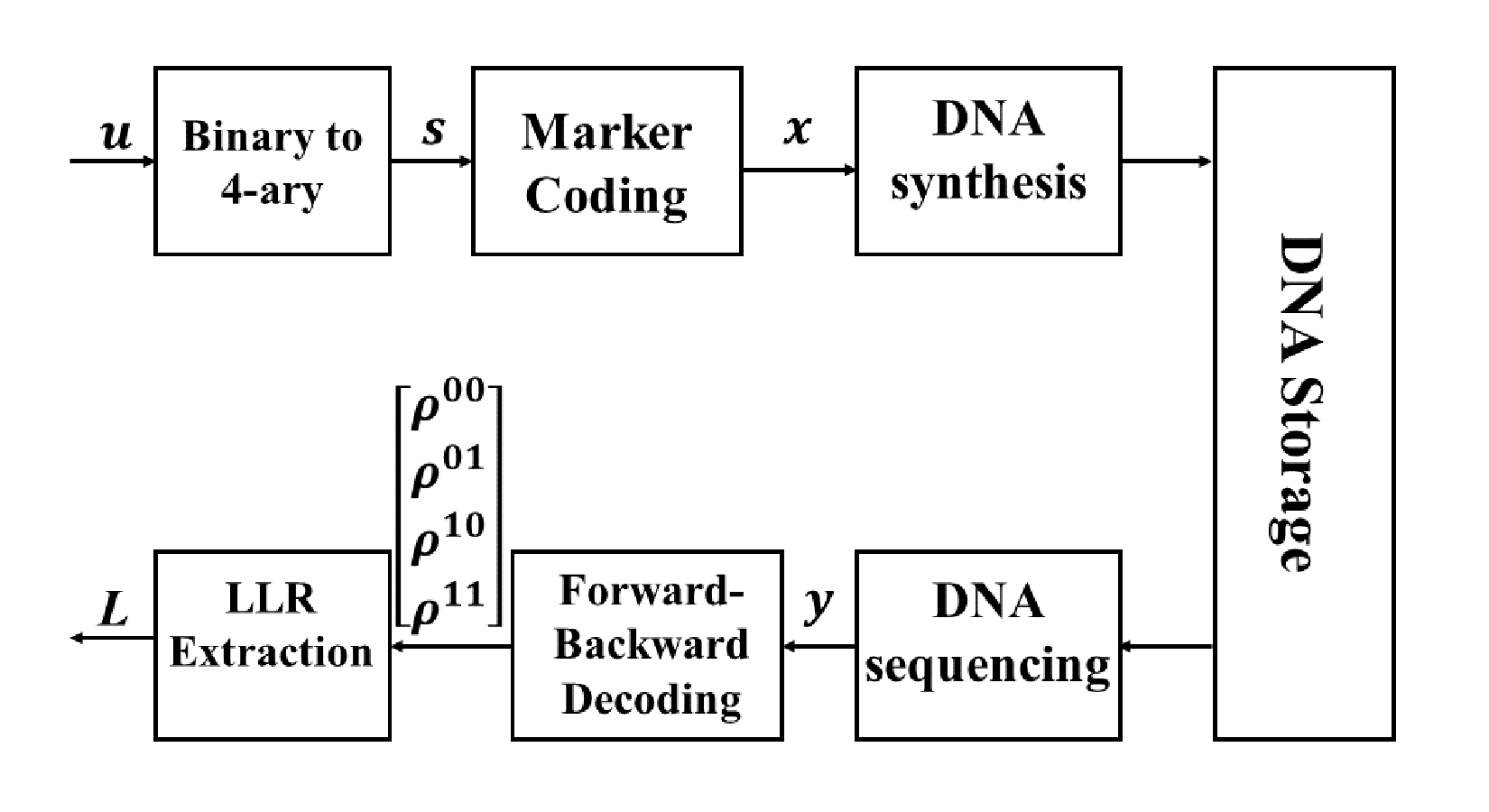}

\caption{\protect\label{fig:System-model}Block diagram of the considered DNA
storage system.}
\end{figure}

The considered system model is shown in Fig. \ref{fig:System-model}.
An $n$-bit information sequence, $\boldsymbol{u}$, is converted
to a $4$-ary vector $\boldsymbol{s}$ of length $\frac{n}{2}$ (assuming
$n$ is even). We denote the $4$-ary alphabet by $\left\{ 00,01,10,11\right\} $
and the $i$-th symbol of $\boldsymbol{s}$ by $s_{i}\in\left\{ 00,01,10,11\right\} $.
The binary to $4$-ary mapping is defined as $s_{i}=u_{2i-1}u_{2i}$,
where $u_{j}\in\left\{ 0,1\right\} $ denotes the $j$-th bit of $\boldsymbol{u}$.
The marker coding process combines known symbols, referred to as marker
symbols, with $\boldsymbol{s}$, in order to protect the data against
synchronization errors. This concatenation leads to a $4$-ary codeword,
$\boldsymbol{x}$. The rate of the marker code, denoted as $r_{M}$,
is determined by dividing the length of $\boldsymbol{s}$ by the length
of $\boldsymbol{x}$.

The codeword, $\boldsymbol{x}$, is synthesized into a DNA strand,
by employing a one-to-one mapping between the $4$-ary alphabet symbols
and the four DNA bases. The DNA strands are then stored in a DNA storage
unit. Accessing the stored data is realized through a DNA sequencing
process which converts the DNA strand back into a $4$-ary vector,
$\boldsymbol{y}$. However, the sequencing process is imperfect due
to the occurrence of insertion, deletion, and substitution errors.
In this work, we primarily focus on deletion and substitution errors,
although insertion errors are also included in our numerical results.
More precisely, we model the end-to-end communication channel between
$\boldsymbol{x}$ and $\boldsymbol{y}$ as a deletion/substitution
channel, as follows. (i) The $i$-th symbol of $\boldsymbol{x}$,
denoted by $x_{i}$, is either deleted with a probability $p_{d}$
or is transmitted with a probability $1-p_{d}$, where a deletion
event is defined as the event that no output symbol is generated corresponding
to the channel input, $x_{i}$. (ii) If $x_{i}$ is not deleted, it
is either correctly received, with probability $1-p_{s}$, or is substituted
by a symbol $y_{j}\neq x_{i}$, where $y_{j}$ is a realization of
a random variable which is uniformly distributed over the set $\left\{ 00,01,10,11\right\} \backslash\left\{ x_{i}\right\} $.
The deletion/substitution events are independent across the entire
sequence. Channel parameters $p_{d}$ and $p_{s}$ are referred to
as the deletion probability and the substitution probability, respectively.

The received vector, $\boldsymbol{y}$, is delivered to an FB decoder
that generates four real-valued (soft) output vectors $\boldsymbol{\rho}^{00},\boldsymbol{\rho}^{01}$,
$\boldsymbol{\rho}^{10},\boldsymbol{\rho}^{11}$. The $i$-th entries
of these vectors, denoted by $\rho_{i}^{00},\ldots,\rho_{i}^{11}$,
are the \textit{a posteriori} probabilities calculated by the FB decoder,
i.e.,

\begin{equation}
\left[\begin{array}{c}
\rho_{i}^{00}\\
\rho_{i}^{01}\\
\rho_{i}^{10}\\
\rho_{i}^{11}
\end{array}\right]=\left[\begin{array}{c}
Pr\left(s_{i}=00|\boldsymbol{y}\right)\\
Pr\left(s_{i}=01|\boldsymbol{y}\right)\\
Pr\left(s_{i}=10|\boldsymbol{y}\right)\\
Pr\left(s_{i}=11|\boldsymbol{y}\right)
\end{array}\right]\label{eq:aposteriori_probs}
\end{equation}

Finally, a vector of log-likelihood ratios (LLRs), $\boldsymbol{L}=\left(L_{1},\ldots,L_{n}\right)$,
corresponding to the binary information sequence, $\boldsymbol{u}$,
is derived by using

\begin{equation}
L_{j}=\log\frac{Pr\left(u_{j}=1|\boldsymbol{y}\right)}{Pr\left(u_{j}=0|\boldsymbol{y}\right)}\label{eq:LLR_def}
\end{equation}
for $j\in\left\{ 1,\ldots,n\right\} $. By recalling the $4$-ary
mapping $s_{i}=u_{2i-1}u_{2i}$ and using (\ref{eq:aposteriori_probs}),
it is straightforward to show that:

\begin{equation}
\begin{array}{cl}
Pr\left(u_{2i-1}=1|\boldsymbol{y}\right) & =\rho_{i}^{10}+\rho_{i}^{11}\\
Pr\left(u_{2i-1}=0|\boldsymbol{y}\right) & =\rho_{i}^{00}+\rho_{i}^{01}\\
Pr\left(u_{2i}=1|\boldsymbol{y}\right) & =\rho_{i}^{01}+\rho_{i}^{11}\\
Pr\left(u_{2i}=0|\boldsymbol{y}\right) & =\rho_{i}^{00}+\rho_{i}^{10}
\end{array}\label{eq:Pu_rho}
\end{equation}
for $i\in\left\{ 1,\ldots,\frac{n}{2}\right\} $. Hence, $L_{j}$
is found by plugging (\ref{eq:Pu_rho}) into (\ref{eq:LLR_def}).
It is worth noting that in most practical schemes, the marker code
is concatenated with an outer error-correcting code, in which case
vector $\boldsymbol{u}$ of Fig. \ref{fig:System-model} is a codeword
of the outer code. The LLR vector $\boldsymbol{L}$ is delivered to
the outer code's decoder, which updates the LLR values and feeds them
back to the inner decoder for subsequent iterations. An estimate of
$\boldsymbol{u}$ is generated after a specified number of LLR exchanges
between the inner and the outer decoders.

\section{\protect\label{sec:Half-Marker-Codes:-Definition}Half-Marker Codes}

In this section, we propose half-marker codes as an alternative to
standard marker codes. We define a half-marker symbol as a symbol
in one of the forms: $\textup{1x, 0x, x1, or x0}$, where $\textup{x}$
denotes an information bit which could be either $0$ or $1$, with
equal probability. For example, a $\textup{1x}$ half-marker symbol
is defined as a symbol that takes on the values $10$ or $11$, each
with probability $0.5$. Using this definition, corresponding to each
standard marker code which periodically inserts $N_{m}$ markers,
we will define a half-marker code that periodically inserts $2N_{m}$
half-markers. 

To formally define the proposed half-marker codes, let us first give
a formal definition for standard marker codes. Let $\boldsymbol{m}=\left(b_{1}b_{1}^{'},\ldots,b_{N_{m}}b_{N_{m}}^{'}\right)$
be a $4$-ary vector of length $N_{m}$ (i.e., $b_{i}\in\left\{ 0,1\right\} $
and $b_{i}^{'}\in\left\{ 0,1\right\} $). Let $N_{p}>N_{m}$ be an
integer, and define $N_{d}=N_{p}-N_{m}$. Define an $\left(N_{p},\boldsymbol{m}\right)$
standard marker code as a code that encodes an input vector $\boldsymbol{s}=\left(s_{1},\ldots,s_{\frac{n}{2}}\right)$
into a codeword $\boldsymbol{x}$ such that:

\begin{equation}
\begin{cases}
\begin{array}{ll}
\boldsymbol{x}_{kN_{p}+1}^{\left(k+1\right)N_{p}}=\left(\boldsymbol{m},\boldsymbol{\boldsymbol{s}}_{kN_{d}+1}^{\left(k+1\right)N_{d}}\right), & k\in\left\{ 0,\ldots,K-1\right\} \\
\boldsymbol{x}_{KN_{p}+1}^{KN_{p}+N_{m}+r}=\left(\boldsymbol{m},\boldsymbol{\boldsymbol{s}}_{KN_{d}+1}^{KN_{d}+r}\right), & r\neq0
\end{array}\end{cases}\label{eq:x_standard_marker_def}
\end{equation}
where $\frac{n}{2}=KN_{d}+r$ for integers $K\geq0$ and $0\leq r\leq N_{d}-1$;
$\boldsymbol{x}_{kN_{p}+1}^{\left(k+1\right)N_{p}}$ denotes the vector
$\left(x_{kN_{p}+1},\ldots,x_{\left(k+1\right)N_{p}}\right)$, and
$\left(\boldsymbol{m},\boldsymbol{s}_{kN_{d}+1}^{\left(k+1\right)N_{d}}\right)$
denotes concatenation of vectors $\boldsymbol{m}$ and $\boldsymbol{\boldsymbol{s}}_{kN_{d}+1}^{\left(k+1\right)N_{d}}$.
In words, the codeword $\boldsymbol{x}$ expressed in (\ref{eq:x_standard_marker_def})
is generated by taking the following steps. First, the encoder calculates
that for transmitting all $\frac{n}{2}$ symbols of $\boldsymbol{s}$,
$K$ segments of length $N_{d}$ and one extra segment of length $r$
is required, where $r\neq0$ if and only if $N_{d}$ does not divide
$\frac{n}{2}$. Then, the encoder periodically inserts the $N_{m}$
marker symbols, specified by the vector $\boldsymbol{m}$, before
each length-$N_{d}$ segment of the data vector $\boldsymbol{s}$.
Finally, if $r\neq0$, the encoder appends one last segment consisting
of the marker symbols followed by the $r$ remaining symbols of $\boldsymbol{s}$.

Now, let us define a vector of $2N_{m}$ half-marker symbols with
marker bits corresponding to $\boldsymbol{m}=\left(b_{1}b_{1}^{'},\ldots,b_{N_{m}}b_{N_{m}}^{'}\right)$,
and information bits corresponding to $\boldsymbol{u}_{2i+1}^{2i+2N_{m}}$,
as follows:

\begin{equation}
\boldsymbol{h}_{2i+1}^{2i+2N_{m}}=\left(b_{1}u_{2i+1},b_{1}^{'}u_{2i+2},\ldots,b_{N_{m}}u_{2N_{m}+1},b_{N_{m}}^{'}u_{2N_{m}+2}\right)\label{eq:h_halfmarker_def}
\end{equation}
From the mapping $s_{j}=u_{2j-1}u_{2j}$, it is clear that the $N_{m}$
information symbols, $\boldsymbol{s}_{i+1}^{i+N_{m}}$, are embedded
in the $2N_{m}$ half-marker symbols, $\boldsymbol{h}_{2i+1}^{2i+2N_{m}}$.
We define an $\left(N_{p},\boldsymbol{m}\right)$ half-marker code,
for $N_{p}\geq2N_{m}$, as a code that generates a codeword $\boldsymbol{x}$
as follows:

\begin{equation}
\begin{cases}
\begin{array}{ll}
\boldsymbol{x}_{kN_{p}+1}^{\left(k+1\right)N_{p}}=\left(\boldsymbol{h}_{2kN_{d}+1}^{2kN_{d}+2N_{m}},\boldsymbol{\boldsymbol{s}}_{kN_{d}+N_{m}+1}^{\left(k+1\right)N_{d}}\right), & 0\leq k<K\\
\boldsymbol{x}_{KN_{p}+1}^{KN_{p}+r}=\boldsymbol{h}_{2KN_{d}+1}^{2KN_{d}+2r}, & 0<r\leq2N_{m}\\
\boldsymbol{x}_{KN_{p}+1}^{KN_{p}+r}=\left(\boldsymbol{h}_{2KN_{d}+1}^{2KN_{d}+2N_{m}},\boldsymbol{\boldsymbol{s}}_{KN_{d}+N_{m}+1}^{KN_{d}+r}\right), & r>2N_{m}
\end{array}\end{cases}\label{eq:x_halfmarker_def}
\end{equation}
The first row in (\ref{eq:x_halfmarker_def}) states that the half-marker
code periodically inserts $2N_{m}$ half-marker symbols before each
block of $N_{d}-N_{m}$ information symbols. Since $N_{m}$ information
symbols, $\boldsymbol{\boldsymbol{s}}_{kN_{d}+1}^{kN_{d}+N_{m}}$,
are embedded into the $2N_{m}$ half-marker symbols, $\boldsymbol{h}_{2kN_{d}+1}^{2kN_{d}+2N_{m}}$,
each $N_{p}$ symbols of the codeword $\boldsymbol{x}$ include a
total of $N_{d}$ information symbols. Therefore, the rate of the
proposed $\left(N_{p},\boldsymbol{m}\right)$ half-marker code is
equal to the rate of the $\left(N_{p},\boldsymbol{m}\right)$ standard
marker code. 

\begin{figure}
\centering\includegraphics[scale=0.25]{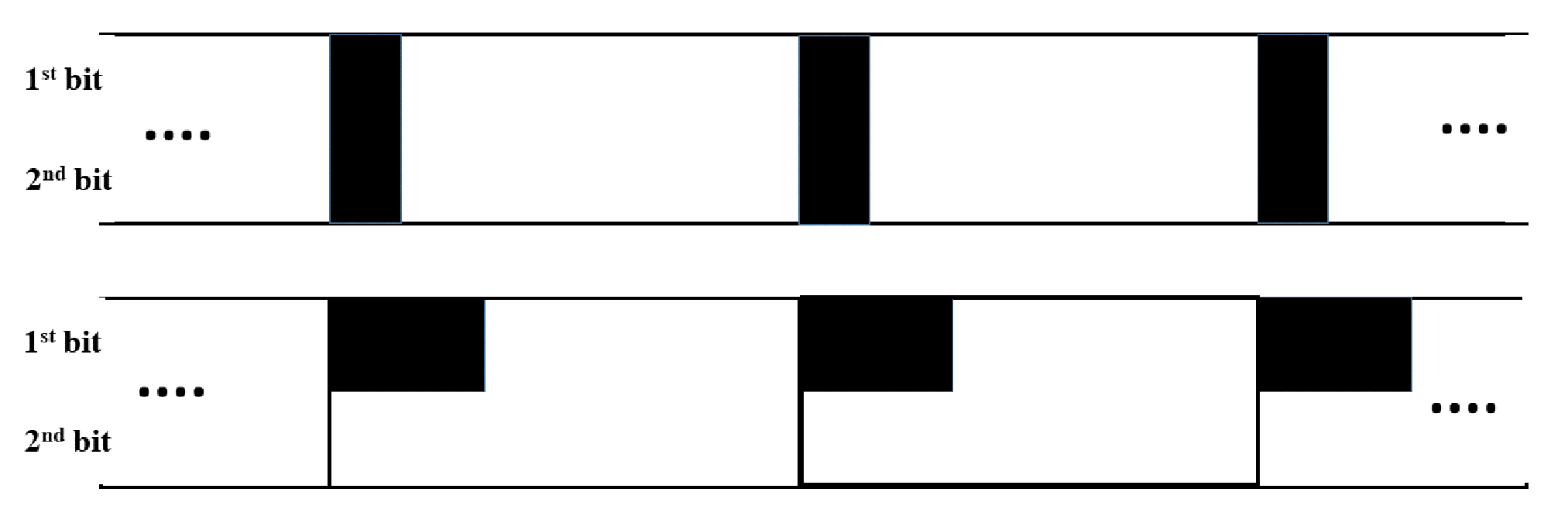}

\caption{\protect\label{fig:standard_half_marker_comparison}Comparison between
inserting standard markers (Top) and half-markers (Bottom) within
a $4$-ary data sequence. Marker bits are shown in black, and data
bits are shown in white.}
\end{figure}

Figure \ref{fig:standard_half_marker_comparison} illustrates and
compares the placement of standard markers and half-markers within
a data sequence. To intuitively explain the potential advantage of
half-marker codes, note that a 4-ary transmission can be interpreted
as simultaneous transmission of two parallel binary sequences (as
also depicted in Fig. \ref{fig:standard_half_marker_comparison}).
Since insertions and deletions introduced by the IDS channel affect
both binary sequences at the same positions, restoring synchronization
in just one of them is theoretically sufficient. Therefore, a logical
strategy is to allocate more markers to one sequence to enhance its
resilience against insertion-deletion noise while omitting markers
from the other---effectively implementing half-markers.

Note that in order to properly decode half-marker codes, the FB decoding
algorithm must be modified to account for the changed statistics of
the transmitted codeword, $\boldsymbol{x}$. A detailed version of
the FB decoding equations is presented in \cite{Haghighat2025}, where
the effect of this modification is reflected in the definition of
the alignment function, $\zeta$, defined as follows:

\begin{equation}
\zeta\left(j,a\right)=\sum_{a'\in\left\{ 00,\ldots,11\right\} }P\left(x_{j}=a'\right)\times f\left(a',a\right),\label{eq:zeta_definition}
\end{equation}
where $f\left(a',a\right)=1-p_{s}$ for $a=a'$ and $f\left(a',a\right)=\frac{p_{s}}{3}$
for $a\neq a'$. When a half-marker symbol is placed at the $j$-th
position, the symbol $x_{j}$ becomes a random variable that takes
two possible values with equal probabilities, i.e., $P\left(x_{j}=a'\right)=0.5$
for two values of $a'$. This differs from the case of a standard
marker symbol, where $P\left(x_{j}=a'\right)=1$ when $a'$ is equal
to the marker symbol, and is zero otherwise. Function $\zeta$ defined
above is further employed to calculate the forward and backward coefficients,
referred to as $\alpha$ and $\beta$ (see \cite{Haghighat2025}-Equations
(15) and (16)). 

To compare the theoretical performance limits of $\left(N_{p},\boldsymbol{m}\right)$
standard and half-marker codes, we employ the \textit{overall achievable
rate} as the theoretical performance metric. Consider the concatenation
of a binary outer code with the marker code, such that the binary
sequence $\boldsymbol{u}$ shown in Fig. \ref{fig:System-model} is
a codeword generated by that binary outer code, and the LLR vector,
$\boldsymbol{L}$, is delivered to a decoder of that binary outer
code to obtain an estimate of $\boldsymbol{u}$. Let $I\left(\boldsymbol{u};\boldsymbol{L}\right)$
denote the mutual information between $\boldsymbol{u}$ and its corresponding
LLR vector, $\boldsymbol{L}$. Then, $\frac{1}{n}I\left(\boldsymbol{u};\boldsymbol{L}\right)$
is an achievable rate for the above-mentioned binary outer code. Hence,
the following overall rate is achievable when a marker code with rate
$r_{M}$ is concatenated with a binary outer code:

\begin{equation}
R_{a}=\frac{r_{M}}{n}\times I\left(\boldsymbol{u};\boldsymbol{L}\right).\label{eq:R_achievable}
\end{equation}

To estimate $I(\boldsymbol{u};\boldsymbol{L})$, we adopt the approach
proposed in \cite{Wang2011}, as explained in the following (note
that, with a slight abuse of notation, random variables and their
realizations are represented using the same symbol). Assuming that
$u_{j}$'s are unbiased, independent, and identically distributed,
we approximate $I(\boldsymbol{u};\boldsymbol{L})\approx\sum_{j=1}^{n}I(u_{j};L_{j}),$where
\begin{equation}
I(u_{j};L_{j})=H(L_{j})-H(L_{j}|u_{j}),\label{eq:mut_info}
\end{equation}
and $H(\cdot)$ denotes entropy. Notably, 
\begin{equation}
H(L_{j}|u_{j})=\frac{1}{2}H(L_{j}|u_{j}=0)+\frac{1}{2}H(L_{j}|u_{j}=1).\label{eq:cond_entropy}
\end{equation}

To estimate $H(L_{j})$, $H(L_{j}|u_{j}=0)$, and $H(L_{j}|u_{j}=1)$,
we approximate their empirical distributions via simulation. Specifically,
we choose a large $n$ (e.g., $10^{4}$), randomly generate $u_{j}$'s,
simulate a realization of the channel, perform forward-backward decoding,
and compute $L_{j}$'s. We then define two sets, $\mathcal{A}_{0}$
and $\mathcal{A}_{1}$, and update them as follows. If $u_{j}=0$,
we append $L_{j}$ to $\mathcal{A}_{0}$; otherwise, we append $L_{j}$
to $\mathcal{A}_{1}$. This process is repeated multiple times (e.g.,
$10000$ iterations). Finally, the recorded values in $\mathcal{A}_{0}$,
$\mathcal{A}_{1}$, and $\mathcal{A}_{0}\cup\mathcal{A}_{1}$ are
used to approximate $H(L_{j}|u_{j}=0)$, $H(L_{j}|u_{j}=1)$, and
$H(L_{j})$, respectively, using histograms. 

In addition to the overall achievable rate, the next section also
compares the performance of half-marker and standard marker codes
using more practical metrics, including bit error and symbol error
rates.

\section{\protect\label{sec:Numerical-Results}Numerical Results}

In this section, we provide numerical results to evaluate the performance
of half-marker codes. We consider two half-marker codes, denoted by
HMC1 and HMC2, corresponding to two standard marker codes, denoted
by SMC1 and SMC2, as follows. SMC1 is an $\left(N_{p},\boldsymbol{m}_{1}\right)$
standard marker code with $\boldsymbol{m}_{1}=\left(10\right)$, i.e.,
a code that inserts a marker symbol, $10$, every $N_{p}$ symbols.
HMC1 is an $\left(N_{p},\boldsymbol{m}_{1}\right)$ half-marker code,
i.e., two half-marker symbols are inserted every $N_{p}$ symbols,
and the inserted half-marker symbols are in the form $\textup{1x,\,0x}$.
SMC2 is an $\left(N_{p},\boldsymbol{m}_{2}\right)$ standard marker
code where two marker symbols, $\boldsymbol{m}_{2}=\left(01,10\right)$,
are inserted every $N_{p}$ symbols; and HMC2 is an $\left(N_{p},\boldsymbol{m}_{2}\right)$
half-marker code where four half-marker symbols are inserted every
$N_{p}$ symbols, and the inserted half-marker symbols are in the
form $\textup{0x,1x,1x,0x}$.

Table \ref{tab:Ra_pd5_ps2} shows the overall achievable rate, $R_{a}$,
defined in (\ref{eq:R_achievable}), for different marker coding schemes.
In order to obtain an accurate estimate of $R_{a}$, we allow a large
information sequence length of $n=10^{4}$ bits. Note that by increasing
$N_{p}$, the marker code rate, $r_{M}$, will increase, whereas $I\left(\boldsymbol{u};\boldsymbol{L}\right)$
will decrease. Due to this trade-off, there exists an optimal value
of $N_{p}$ for which $R_{a}=\frac{r_{M}}{n}\times I\left(\boldsymbol{u};\boldsymbol{L}\right)$
is maximized. As highlighted in Table \ref{tab:Ra_pd5_ps2}, the maximum
overall achievable rate is $.53$, $.59$, $.56$, and $.60$, for
SMC1, HMC1, SMC2, and HMC2, respectively. It is observed that the
half-marker codes have the potential to increase the maximum achievable
rates compared to their corresponding standard marker codes. For instance,
the maximum achievable rate offered by HMC1 is $11\%$ greater than
that offered by SMC1. Also, we observe that for any given $N_{p}$,
the value of $R_{a}$ is larger for HMC1 (HMC2) compared to SMC1 (SMC2).
Since for a given $N_{p}$, the rates of HMC1 and SMC1 are equal,
we conclude that HMC1 achieves a larger mutual information, $I\left(\boldsymbol{u};\boldsymbol{L}\right)$,
compared to SMC1. Similarly, for any given $N_{p}$, HMC2 achieves
a larger mutual information compared to SMC2.

\begin{table}
\caption{\protect\label{tab:Ra_pd5_ps2}Overall achievable rate, $R_{a}$,
when $p_{d}=0.05$ and $p_{s}=0.02$. }

\centering%
\begin{tabular}{|c|c|c|c|c|}
\hline 
$N_{p}$  & SMC1  & HMC1  & SMC2  & HMC2\tabularnewline
\hline 
\hline 
$5$  & $.48$  & $.55$  & $.43$  & $.50$\tabularnewline
\hline 
$7$  & $.52$  & $.58$  & $.54$  & $.55$\tabularnewline
\hline 
$9$  & \textbf{$\boldsymbol{.53}$}  & $.\boldsymbol{59}$  & $.55$  & $.59$\tabularnewline
\hline 
$11$  & $.51$  & $.53$  & $\boldsymbol{.56}$  & $\boldsymbol{.60}$\tabularnewline
\hline 
$13$  & $.48$  & $.51$  & $.56$  & $.60$\tabularnewline
\hline 
$15$  & $.46$  & $.50$  & $.55$  & $.58$\tabularnewline
\hline 
\end{tabular}
\end{table}

\begin{figure}
\centering\includegraphics[scale=0.35]{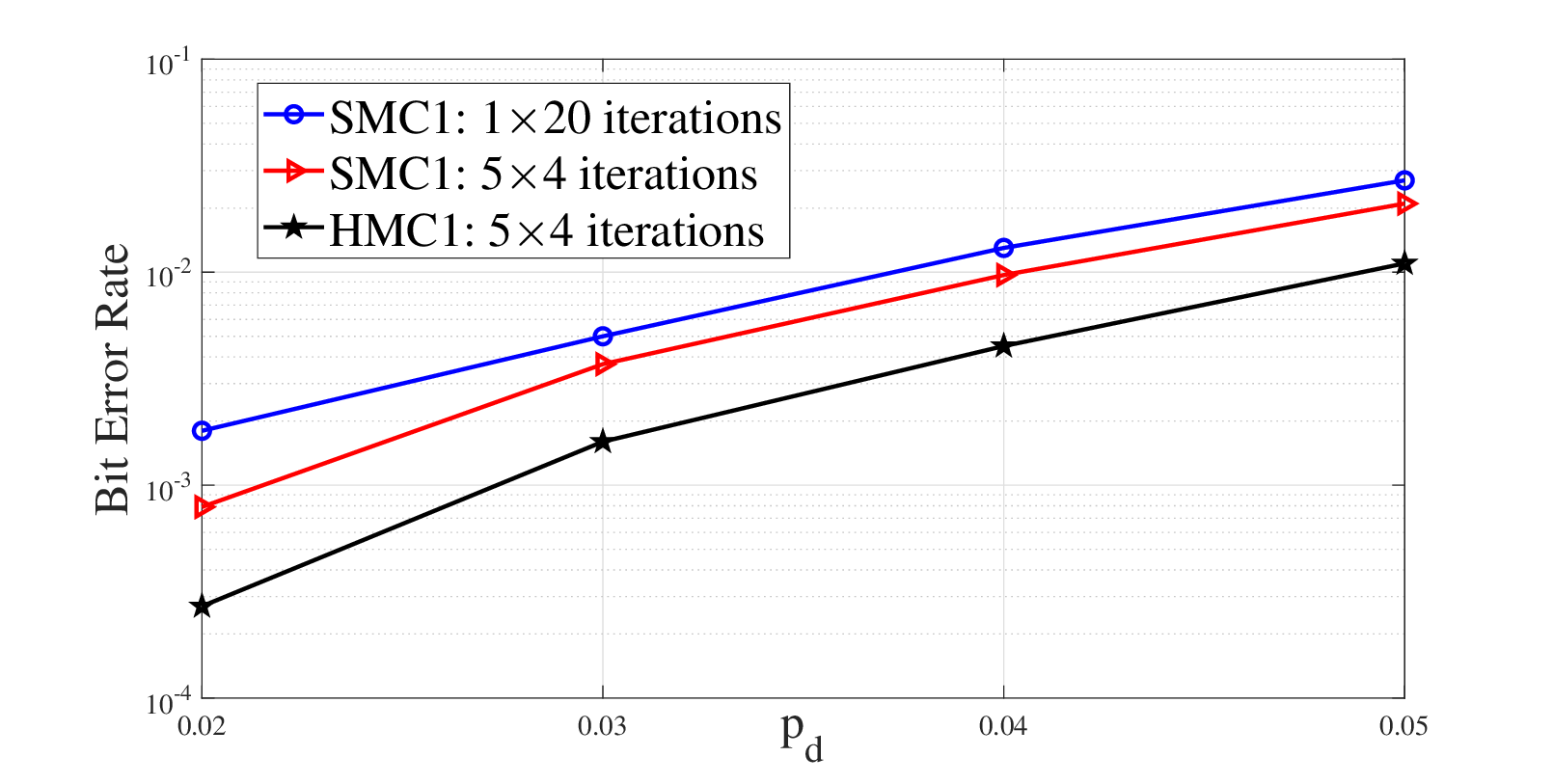}

\caption{\protect\label{fig:Bit-error-rates}Bit error rates of concatenated
LDPC-marker coding schemes, when $p_{s}=0.02$ and $N_{p}=6$. }
\end{figure}

In Fig. \ref{fig:Bit-error-rates}, we simulate a concatenated LDPC-marker
coding scheme, where a $\left(300,150\right)$ LDPC code is employed
as the outer code, i.e., vector $\boldsymbol{u}$ of Fig. \ref{fig:System-model}
is a codeword of this LDPC code. The employed LDPC code is regular
and is constructed using Gallager's approach \cite{Gallager1962}
with a variable node degree of $3$ and a check node degree of $6$.
At the decoder side, the LLR vector, $\boldsymbol{L}$, generated
by the FB decoder, is passed to an LDPC decoder which updates these
LLRs and returns them as \textit{a priori} information to the FB decoder.
This exchange continues for a specified number of iterations, after
which a hard decision, denoted as $\hat{\boldsymbol{u}}$, is made
using the most up-to-date LLR values. We define the end-to-end bit
error rate (BER) as $E\left[\frac{1}{n}\sum_{j=1}^{n}\left(u_{j}\oplus\hat{u}_{j}\right)\right]$,
where $E\left[.\right]$ and $\oplus$ denote statistical expectation
and exclusive-or (XOR) operation, respectively. We plot this BER when
SMC1 or HMC1 is employed as the inner code. The label ``$5\times4$
iterations'' in Fig. \ref{fig:Bit-error-rates} indicates that each
FB decoding step is followed by four LDPC decoding iterations, and
the information is exchanged between the two decoders five times.
It is observed that the scheme which employs HMC1 significantly outperforms
the one which employs SMC1. Therefore, half-marker codes may significantly
improve performance compared to standard marker codes, even at short
block lengths, $n$. This observation is important since the state
of the art DNA storage systems are only capable of synthesizing short-length
DNA strands (due to the current synthesis technology limitations \cite{Milenkovic2024}).
Also, a result labeled ``$1\times20$ iterations,'' corresponding
to an FB decoding step followed by $20$ LDPC decoding iterations
without exchanging information between the two decoders, is included
in Fig. \ref{fig:Bit-error-rates}. 

\begin{figure}
\centering\includegraphics[scale=0.35]{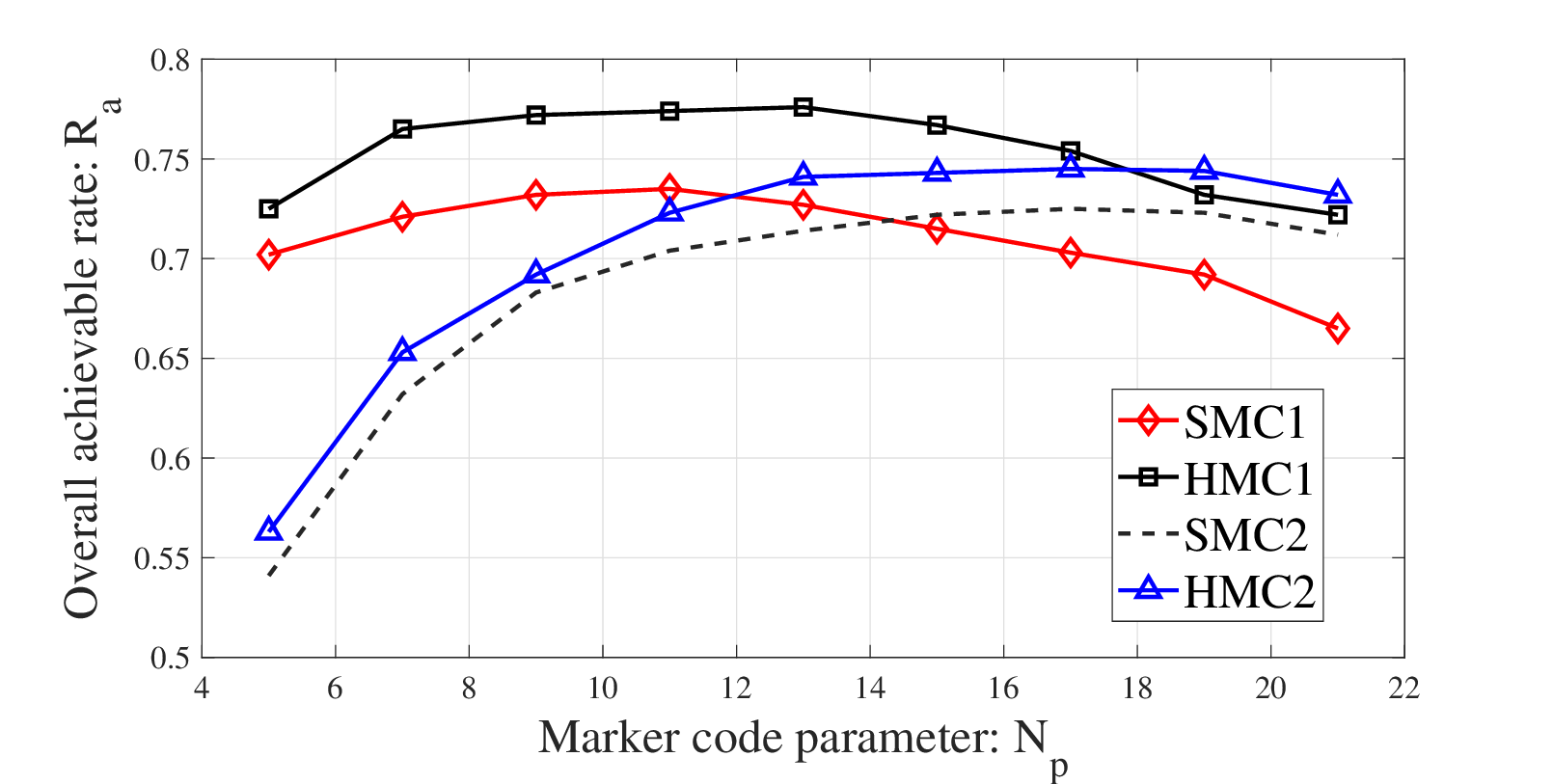}

\caption{\protect\label{fig:Achievable-rates-for}Overall achievable rate,
for $p_{d}=0.02$ and $p_{s}=0.01$.}
\end{figure}

Figure \ref{fig:Achievable-rates-for} depicts the achievable rates
for different marker patterns, determined via Monte Carlo simulation
as described in detail in Section \ref{sec:Half-Marker-Codes:-Definition}.
It is observed that the maximum achievable rate is larger in the case
of half-marker codes. Specifically, the maximum achievable rates for
SMC1 and SMC2 are $0.736$ (achieved at $N_{p}=13$) and $0.725$
(achieved at $N_{p}=17$), respectively; whereas the maximum achievable
rates for HMC1 and HMC2 are $0.776$ (achieved at $N_{p}=13$) and
$0.745$ (achieved at $N_{p}=17$), respectively. It is also observed
that the half-marker code HMC1 outperforms the other codes in this
setup for all values of $5\leq N_{p}\leq17$.

\begin{figure}
\centering\includegraphics[scale=0.35]{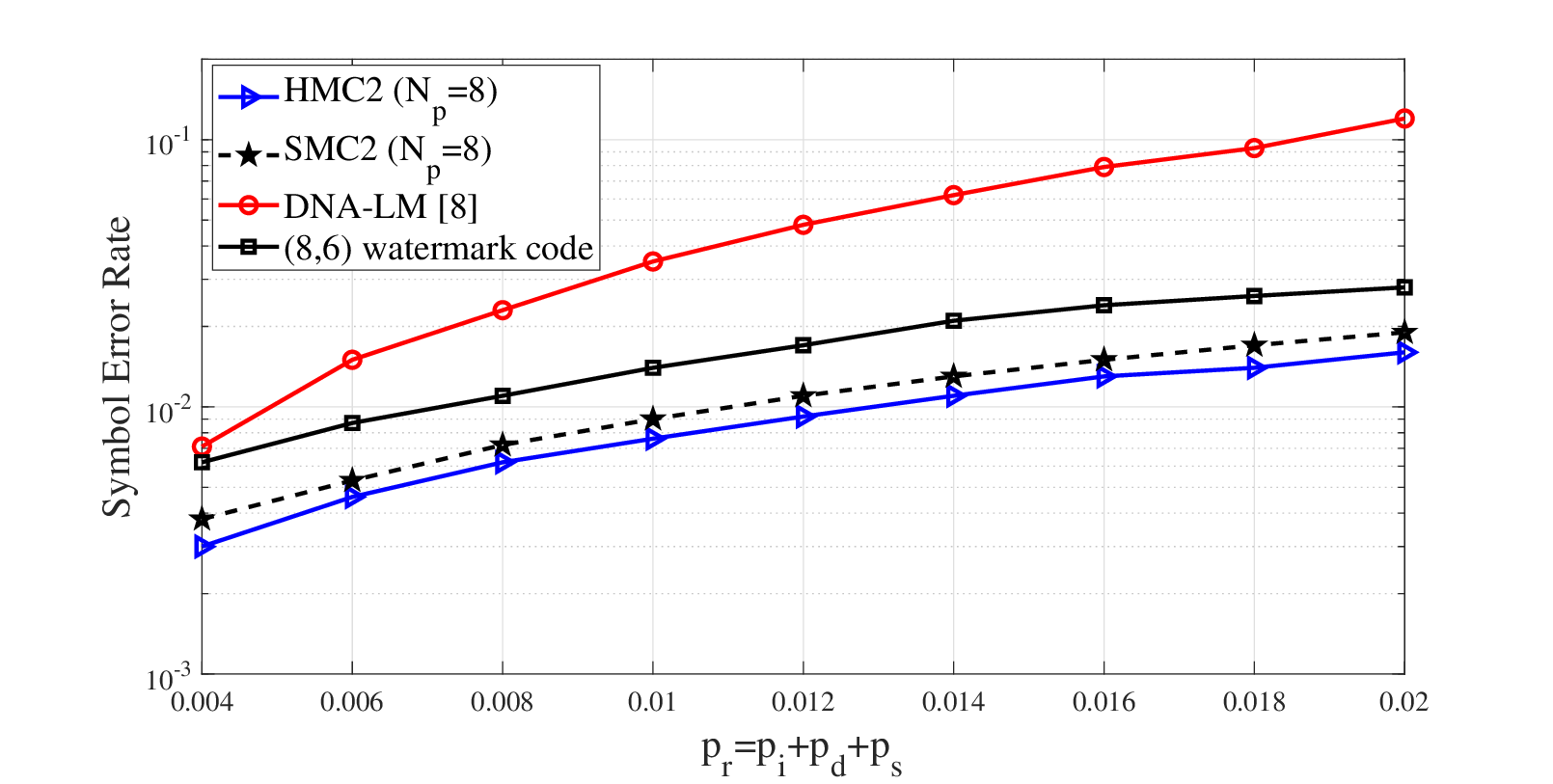}

\caption{\protect\label{fig:schemes_comparison}Comparison between the symbol
error rates achieved by different synchronization codes.}
\end{figure}

Figure \ref{fig:schemes_comparison} compares the symbol error rates
(SER) of different coding schemes. To evaluate the SER, a hard decision
on the symbol sequence is made after FB decoding, as follows: 

\begin{equation}
\hat{s}_{i}=\underset{bb'\in\left\{ 00,\ldots,11\right\} }{\textup{argmax}}\rho_{i}^{bb'},\,i\in\left\{ 1,\ldots,\frac{n}{2}\right\} \label{eq:s_hat_decision}
\end{equation}
The SER is then computed via Monte Carlo simulation of $E\left[\frac{2}{n}\sum_{i=1}^{n/2}\boldsymbol{1}_{s_{i}\neq\hat{s}_{i}}\right]$
where $\boldsymbol{1}_{s_{i}\neq\hat{s}_{i}}$ is an indicator function
that returns $1$ if $s_{i}\neq\hat{s}_{i}$ and $0$ otherwise. The
watermark code is implemented as described in \cite{Davey2001}, while
the DNA-LM code follows \cite{Yan2023}, with its SER curve shown
in (\cite{Yan2023}, Fig. 6-a). The comparisons consider insertion
errors using the IDS channel model from (\cite{Yan2023}, Fig. 2).
Here, $p_{i}$ denotes the insertion probability, and $p_{r}=p_{i}+p_{d}+p_{s}$
represents the mutation error rate \cite{Yan2023}, where we assume
$p_{s}=2p_{i}=2p_{d}$ as in \cite{Yan2023}. For the DNA-LM code,
the original data sequence $\boldsymbol{s}$ consists of $180$ symbols
(\cite{Yan2023}, Table 4), and the encoded sequence $\boldsymbol{x}$
has $246$ symbols. For SMC2 and HMC2, $\boldsymbol{s}$ also consists
of $180$ symbols, but it is encoded into $\boldsymbol{x}$ with $240$
symbols by inserting either $2$ standard marker symbols (for SMC2)
or $4$ half-marker symbols (for HMC2) every $N_{p}=8$ symbols. For
the watermark code, a binary sequence $\boldsymbol{u}$ of length
$360$ bits is encoded into $\frac{8}{6}\times360=480$ bits using
an $\left(8,6\right)$ watermark code. These $480$ bits are then
mapped to a symbol sequence $\boldsymbol{x}$ of length $240$. Note
that the DNA-LM code has a slightly longer block length; consequently,
a slightly lower rate compared to the other codes.

As shown in Fig. \ref{fig:schemes_comparison}, the marker codes outperform
both the watermark and DNA-LM codes, with HMC2 achieving the lowest
SER values among all schemes. However, we note that the DNA-LM code
has lower decoding complexity compared to the other schemes \cite{Yan2023}. 

\section{\protect\label{sec:Conclusions}Conclusions}

We propose a new class of marker codes, referred to as half-marker
codes, for DNA data storage systems. Although it cannot be guaranteed
that half-marker codes will always outperform their corresponding
standard marker codes, they provide us with a larger class of marker
codes to select from, and our extensive experiments show that they
have the potential to improve the performance of various system setups
and channel parameters, including the ones that are studied in this
letter. Note that design of optimized concatenated LDPC--half-marker
codes is also possible as a further extension.

\end{document}